\documentclass[DIV=calc, paper=a4, fontsize=11pt]{scrartcl}	 
\usepackage[english]{babel} 
\usepackage{xspace}
\usepackage{bm}
\usepackage[protrusion=true,expansion=true]{microtype} 
\usepackage{amsmath,amsfonts,amsthm, amssymb} 
\usepackage[svgnames]{xcolor} 
\usepackage[hang, small,labelfont=bf,up,textfont=it,up]{caption} 
\usepackage{booktabs} 
\usepackage{fix-cm}	 
\usepackage[pdftex]{graphicx}
\usepackage{hyperref}
\usepackage{sectsty} 
\allsectionsfont{\usefont{OT1}{phv}{b}{n}} 
\usepackage{wasysym}
\usepackage{fancyhdr} 
\usepackage{lastpage} 
\usepackage{setspace}
\usepackage[separate-uncertainty = true, exponent-product = \times]{siunitx}
\usepackage{lmodern}
\usepackage{lettrine} 

\usepackage{titling} 

\newcommand{\HorRule}{\color{DarkGoldenrod} \rule{\linewidth}{1pt}} 

\pretitle{\vspace{-30pt} \begin{flushleft} \HorRule \fontsize{20}{20} \usefont{OT1}{phv}{b}{n} \color{DarkRed} \selectfont} 

\title{Observation of the spin Nernst effect}

\posttitle{\par\end{flushleft}\vskip 0.5em} 

\preauthor{\begin{flushleft}\large \lineskip 0.5em \usefont{OT1}{phv}{b}{sl} \color{DarkRed}} 

\author{S. Meyer$^{1,2}$, Y.-T. Chen$^{3, 4}$, S. Wimmer$^{5}$, M. Althammer$^{1}$, T. Wimmer$^{1,2}$, R. Schlitz$^{1,6,7}$, S. Gepr\"ags$^{1}$, H. Huebl$^{1,2,8}$, D. K\"odderitzsch$^{5}$, H. Ebert$^{5}$, G.E.W. Bauer$^{3,9,10}$, R. Gross$^{1,2,8}$ and S.T.B. Goennenwein$^{1,2,6,7,8}$}

\postauthor{\footnotesize \usefont{OT1}{phv}{m}{sl} \color{Black} 
\begin{enumerate}
	\item {Walther-Mei{\ss}ner-Institut, Bayerische Akademie der Wissenschaften, Walther-Mei{\ss}ner-Stra{\ss}e 8, 85748 Garching, Germany}
    \item {Physik-Department, Technische Universit\"at M\"unchen, 85748 Garching, Germany}
	\item {Kavli Institute of NanoScience, Delft University of Technology, Lorentzweg 1, 2628 CJ Delft, The Netherlands}
	\item {RIKEN Center for Emergent Matter Science (CEMS), 2-1 Hirosawa, Wako, Saitama 351-0198, Japan}
	\item {Department Chemie, Physikalische Chemie, Universit\"at M\"unchen, Butenandtstra{\ss}e 5-13, 81377 M\"unchen, Germany}
    \item {present address: Institut f\"ur Festk\"orperphysik, Technische Universit\"at Dresden, 01062 Dresden, Germany}
    \item {present address: Center for Transport and Devices of Emergent Materials, Technische Universit\"at Dresden, 01062 Dresden, Germany}
    \item {Nanosystems Initiative Munich (NIM), Schellingstra{\ss}e 4, 80799 M\"unchen, Germany}
	\item {Institute for Materials Research, Tohoku University, Sendai, Miyagi 980-8577, Japan}
	\item {WPI Advanced Institute for Materials Research, Tohoku University, Sendai 980-8577, Japan}

\end{enumerate}
\par\end{flushleft}\HorRule} 

\date{\today} 

\begin{document}

\maketitle 

\thispagestyle{fancy} 

\newpage

\textbf{
The observation of the spin Hall effect \cite{Spin-currents:Kato:Science:2004,Wunderlich2005, SHE-transistor:Wunderlich:Science2010} triggered intense research on pure spin current transport \cite{SHE:Sinova:RMP2015}.
With the spin Hall effect \cite{Spin-currents:Kato:Science:2004,Wunderlich2005,Valenzuela:2006,Spin-pumping:saitoh:APL:2006}, the spin Seebeck effect \cite{Caloritronix:SpinSeebeck:Uchida:Saitoh:Nature:2008,Caloritronix:SpinSeebeck:GaMnAs:Jaworski:Myers:NatMat:2010,Longitudinal-spin-Seebeck:Uchida:APL:2010:172505}, and the spin Peltier effect \cite{Flipse_nano_2012,spin-Peltier:Flipse:PhysRevLett.113.027601} already observed, our picture of pure spin current transport is almost complete.
The only missing piece is the spin Nernst (-Ettingshausen) effect, that so far has only been discussed on theoretical grounds \cite{SNE:Cheng:PRB2008,SNE:Liu:SSC2010,Tauber:Gradhand:SNE:PRL:2011,SNE:Wimmer:PRB2013}.
Here, we report the observation of the spin Nernst effect. By applying a longitudinal temperature gradient, we generate a pure transverse spin current in a Pt thin film. For readout, we exploit the magnetization-orientation-dependent spin transfer to an adjacent yttrium iron garnet layer, converting the spin Nernst current in Pt into a controlled change of the longitudinal and transverse thermopower voltage.
Our experiments show that the spin Nernst and the spin Hall effect in Pt are of comparable magnitude, but differ in sign, as corroborated by first-principles calculations.}

\section{Main letter}
\begin{figure}[htbp]
	\centering
	\includegraphics[width=1.00\textwidth]{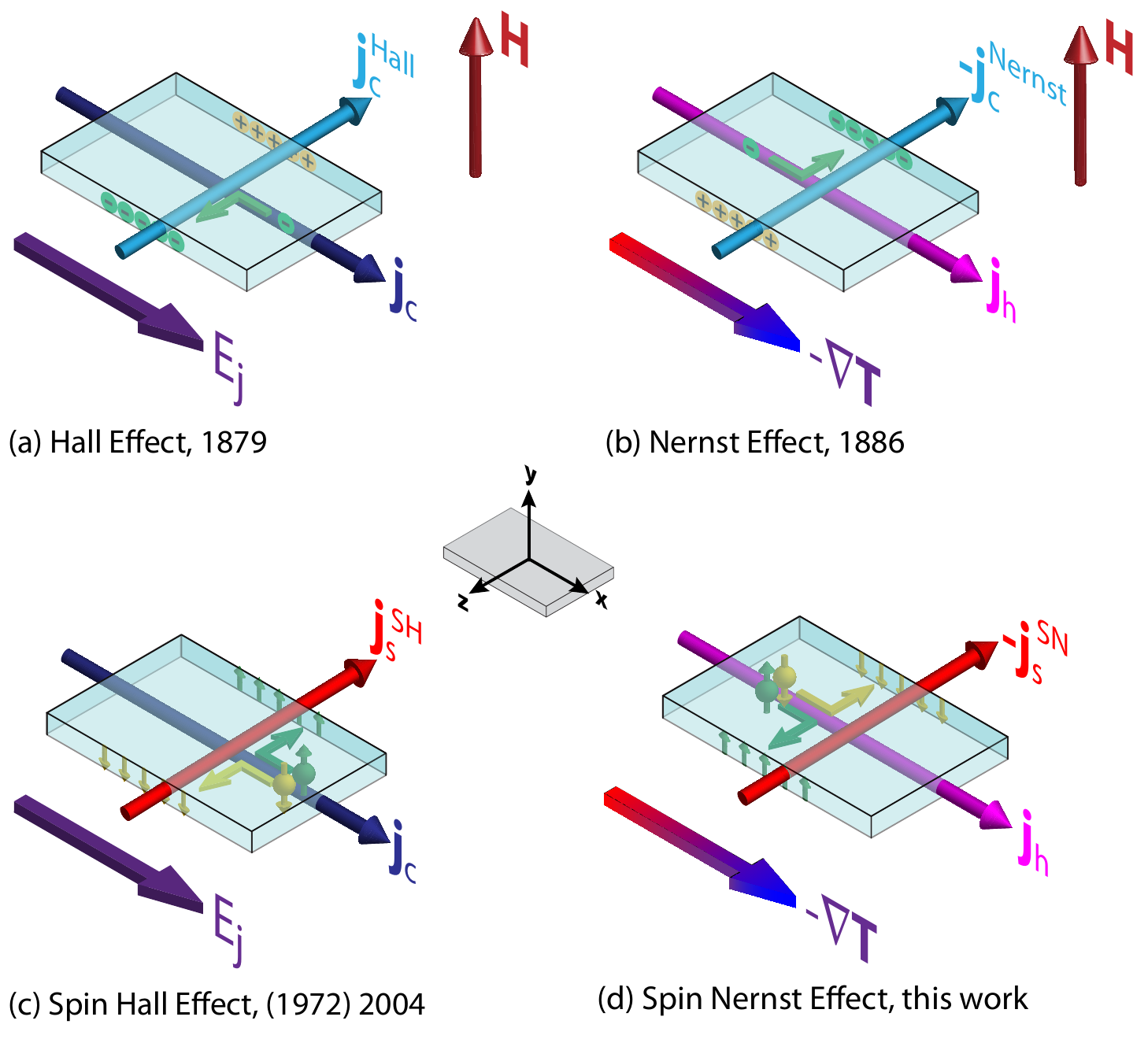}
	\caption{\textbf{Charge and spin-related electric and thermal effects}. \textbf{(a)} In the Hall effect, a transverse charge current density $\mathbf {j}_{\mathrm{c}}^\mathrm{Hall}$ arises when a magnetic field $\mathbf{H}$ and a charge current density $\mathbf {j}_{\mathrm{c}}$ are applied normal to each other. \textbf{(b)} The Nernst effect is the thermal analogue of the Hall effect. The electric effects are shown for negative charge carriers (electrons), translating into negative Hall and Nernst angles. \textbf{(c)} In the spin Hall effect, a transverse spin current density $\mathbf {j}_{\mathrm{s}}^\mathrm{SH}$ perpendicular to the charge current density $\mathbf {j}_{\mathrm{c}}$ is generated due to spin orbit coupling. \textbf{(d)} A transverse spin current density $\mathbf {j}_{\mathrm{s}}^\mathrm{SN}$ is also generated by a longitudinal temperature gradient. This effect has been named \textit{spin Nernst effect} and is experimentally demonstrated here. The spin effects are shown for negative spin Hall and spin Nernst angles. }
	\label{fig1}
\end{figure}

Transverse transport is a key aspect of charge and/or spin motion in the solid state.
In the charge channel, the Hall effect \cite{Hall:original-article} and the Nernst effect \cite{Nernst:Nernst-effect:AnnPhysChem1886} sketched in Fig.~\ref{fig1} (a), (b) enshrine transverse charge motion due to a gradient in the longitudinal potential imposed by an electric or thermal stimulus, respectively.
Since the magnitude of the Hall charge current $\mathbf{j}_{\mathrm{c}}^{\mathrm{Hall}}\propto \theta_{\mathrm{H}}\, \mathbf{j_{\mathrm{c}}} \times \mathbf{H}$ (parameterized by the Hall angle $\theta_\mathrm{H}$, the applied charge current $\mathbf{j_{\mathrm{c}}}$ and the external magnetic field $\mathbf{H}$) is governed by the density of mobile charge carriers in simple metals and semiconductors, Hall effect experiments quickly became a standard tool for material characterization.
As sketched in Fig.~\ref{fig1}(b), the transverse Nernst charge current $\mathbf{j}_{\mathrm{c}}^{\mathrm{Nernst}}\propto \theta_{\mathrm{N}}\nabla T \times \mathbf{H}$ is driven by a temperature gradient $\nabla T$ or the corresponding heat current $\mathbf {j}_\mathrm{h} = -\kappa \nabla T$, where $\kappa$ is the thermal conductivity and $\theta_{\mathrm{N}}$ the Nernst angle.

While first experiments in the spirit of the spin Hall effect have been conducted in the 1970s\,\cite{Chazalviel1972}
, only recently, electrically driven transverse \textit{spin} transport in the form of the spin Hall effect (SHE) \cite{SHE:Dyakonov:Perel:JETPLett:1971,SHE:Hirsch:PRL:1999} resulted in a new paradigm for spin-electronic device design \cite{SHE-transistor:Wunderlich:Science2010, Spin-transistor:Betthausen:Science2012,SHE:Sinova:RMP2015}.
The SHE refers to a transverse pure spin current $\mathbf{j}_{\mathrm{s}}^{\mathrm{SH}}\propto \theta_{\mathrm{SH}} \, \mathbf{j}_\mathrm{c} \times \mathbf{s}$ driven by a charge current density $\mathbf{j}_\mathrm{c}$, see Fig.~\ref{fig1}(c). The spin Hall angle $\theta_{\mathrm{SH}}$ characterizes the charge-to-spin conversion efficiency \cite{SHE:Sinova:RMP2015,SHE-metals:review:Hoffmann:IEEETM2013}.
Analogous to the Nernst effect, the spin Nernst effect (SNE) describes a transverse pure spin current $\mathbf{j}_{\mathrm{s}}^{\mathrm{SN}}\propto \theta_{\mathrm{SN}} \,  \mathbf {j}_\mathrm{h} \times \mathbf{s}$ arising from a longitudinal temperature gradient, cf.~Fig.~\ref{fig1}(d). Here, $\theta_{\mathrm{SN}}$ is the spin Nernst angle \cite{SNE:Cheng:PRB2008,SNE:Liu:SSC2010,Tauber:Gradhand:SNE:PRL:2011,SNE:Wimmer:PRB2013}.
In linear response and Sommerfeld approximation:
\begin{equation}
\left (
\begin{array}{c}\mathbf {j}_{\mathrm{c}}\\ \mathbf j_\mathrm{h} \\ \mathbf {j}_{\mathrm{s,i}}
\end{array} \right ) = \sigma \left ( \begin{array}{ccc}1 & ST & \theta_{\mathrm{SH}} \mathbf{i} \times \\
                                                ST & L_0 T^2 & ST \theta_{\mathrm{SN}} \mathbf{i} \times \\
\theta_{\mathrm{SH}} \mathbf{i} \times & ST \theta_{\mathrm{SN}} \mathbf{i} \times & 1 \end{array} \right )
\left (
\begin{array}{c}\nabla \mu_0 /e\\ -\nabla T /T \\ \nabla \boldsymbol \mu_{\mathrm{si}}/(2e)
\end{array} \right ),
\label{eq:response}
\end{equation}
where the gradients of the electrochemical potential $\mu_0$, $T$ and spin accumulation $\boldsymbol \mu_{\mathrm si}$ are connected via a tensor to $\mathbf{j}_{\mathrm{c}}$, $\mathbf {j}_\mathrm{h}$ and the pure spin current $\mathbf{j}_{\mathrm{s,i}}$ (with spin polarization $\mathbf{s}$ and $i \in \{\mathrm{x,y,z}\}$). The response tensor contains the electrical conductivity $\sigma$, the Seebeck
coefficient $S$, the Lorenz constant $L_0$ and the spin Hall (Nernst) angle $\theta_{\mathrm{SH}}\,\,(\theta_{\mathrm{SN}})$ (for more details see SI).
In spite of its fundamental importance for the understanding of pure spin current transport, the SNE has remained a theoretical conjecture.

\begin{figure}[hp]
	\centering
          \includegraphics[width=\textwidth]{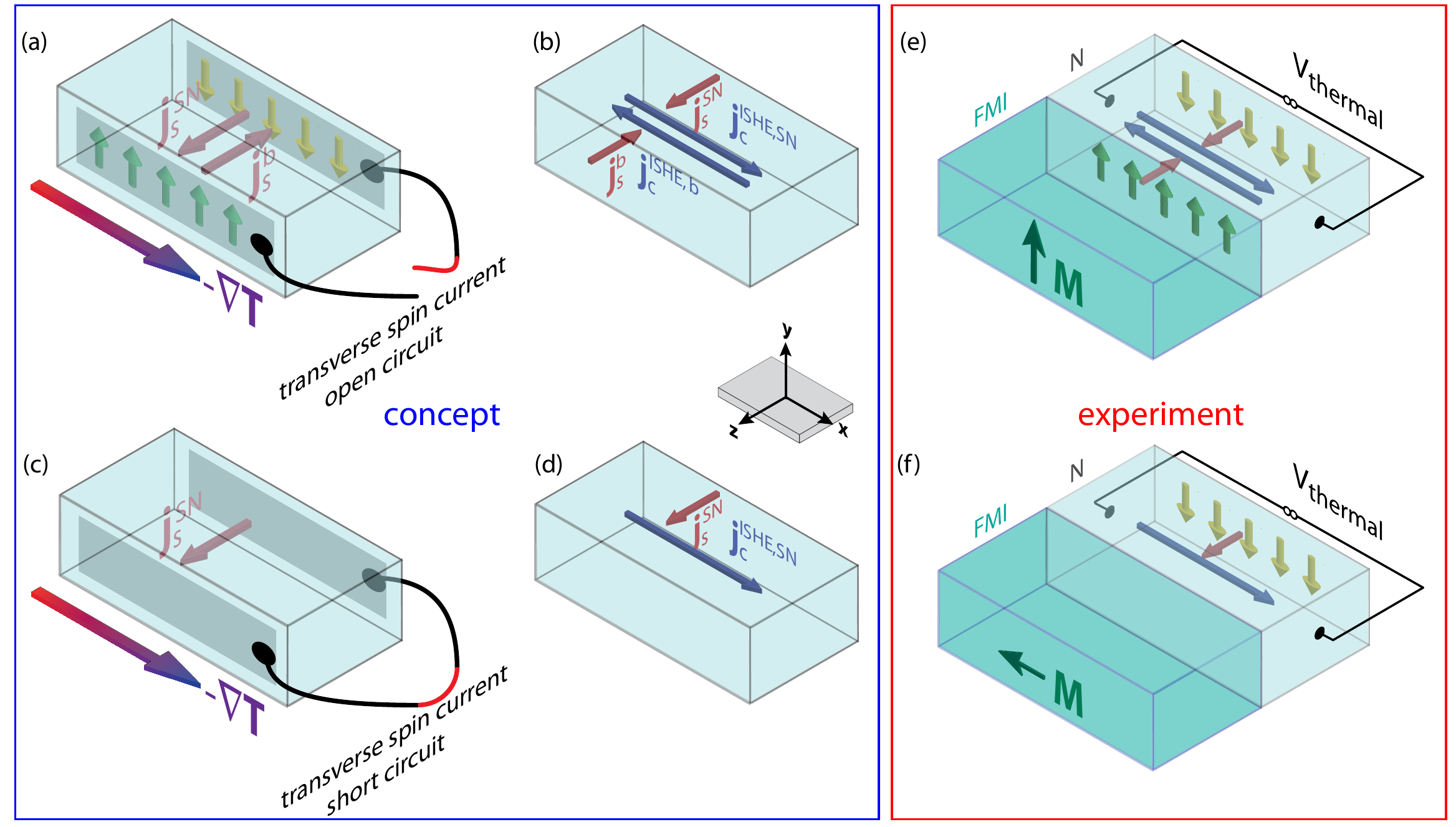}
		\caption{\textbf{Boundary conditions for the spin Nernst magneto-thermopower (SMT)}:
\textbf{(a)}: A temperature gradient $\nabla T$ along $-\mathbf x$ evokes a spin current density $\mathbf {j}_{\mathrm{s}}^\mathrm{SN}$ along $\mathbf z$, leading to a spin-dependent chemical potential along $\mathbf z$. Open circuit boundary conditions depicted in \textbf{(a)} block the transverse spin current, generating a spin current back flow $\mathbf {j}_{\mathrm{s}}^\mathrm{b} = -\mathbf {j}_{\mathrm{s}}^\mathrm{SN}$.
\textbf{(b)} Both spin current densities $\mathbf {j}_{\mathrm{s}}^\mathrm{SN}$ and $\mathbf {j}_{\mathrm{s}}^\mathrm{b}$ give rise to charge current densities $\mathbf {j}_{\mathrm{c}}^\mathrm{ISHE,SN}$ and $\mathbf {j}_{\mathrm{c}}^\mathrm{ISHE,b}$ parallel and antiparallel to $\mathbf x$.
\textbf{(c)} Short-circuiting the spin transport along $\mathbf z$ suppresses the spin-dependent chemical potential and $\mathbf {j}_{\mathrm{s}}^\mathrm{back}$.
\textbf{(d)} The absence of $\mathbf {j}_{\mathrm{s}}^\mathrm{back}$ enhances the net charge current.
\textbf{(e,f)} We utilize an insulating ferrimagnet (FMI) attached to the metal layer (N) to switch between open (no spin transfer torque, panel \textbf{(e)}) and short-circuit (finite spin transfer torque, panel \textbf{(f)}) boundary conditions by the FMI magnetization orientation $\mathbf M$.}
	\label{fig2}
\end{figure}

In this Letter, we report direct experimental evidence for the spin Nernst effect in platinum. In order to quantify the spin Nernst spin current, we modulate the transverse spin current transport boundary conditions and detect the spin accumulation induced by the spin Nernst effect (SNE) in the charge channel, via the inverse spin Hall effect (ISHE)\,\cite{Spin-pumping:saitoh:APL:2006}. In model calculations, we show that the combined action of SNE and ISHE results in a thermopower along (and normal to) the applied temperature gradient. This spin Nernst magneto-thermopower (SMT) is present in any electrical conductor with spin orbit coupling, but usually cannot be discerned from the conventional Seebeck effect, since it has the same symmetry. However, by selectively changing the spin transport boundary conditions, the SMT can be quantitatively extracted and analyzed.\\
The concept is illustrated in Fig.~\ref{fig2}(a-d). A paramagnetic metal film is exposed to a temperature gradient $\nabla T || -\mathbf{x}$. Through the Seebeck effect, a thermopower arises along $\mathbf{x}$. Furthermore, because of the SNE, a transverse spin current $\mathbf{j}_{\mathrm{s}}^{\mathrm{SN}}$ is flowing along $\mathbf{z}$ with spin polarization along $\mathbf{y}$, resulting in a spin accumulation at the metal film boundaries, as sketched in Fig.~\ref{fig2}(a).
The ensuing spin accumulation in turn drives a diffusive spin current $\mathbf{j}_{\mathrm{s}}^{\mathrm{b}}$.
In the steady state, the spin current back flow $\mathbf{j}_{\mathrm{s}}^{\mathrm{b}}=-\mathbf{j}_{\mathrm{s}}^{\mathrm{SN}}$ exactly balances the spin Nernst spin current, such that the net transverse spin current flow vanishes.
Through the ISHE, both $\mathbf{j}_{\mathrm{s}}^{\mathrm{b}}$ and $\mathbf{j}_{\mathrm{s}}^{\mathrm{SN}}$ are accompanied by inverse spin Hall charge currents (cf. Fig.~\ref{fig2}(b)). Since the latter are of equal magnitude but opposite in sign, they  cancel. In that case the charge current vanishes and thereby the SMT.
In contrast, when the transverse spin transport is short-circuited, the spins can not accumulate at the interface [Fig.~\ref{fig2}(c)], such that $\mathbf{j}_{\mathrm{s}}^{\mathrm{b}}$ is suppressed.
As a consequence, only $\mathbf{j}_{\mathrm{s}}^{\mathrm{SN}}$ drives an ISHE charge current that leads to a net
charge current $\mathbf{j}_{\mathrm{c}}^{\mathrm{ISHE, SN}}$ along $\mathbf{x}$, i.e., along the direction of the thermal bias [cf. Fig.~\ref{fig2}(d)].
The combination of spin Nernst and inverse spin Hall effects thereby induces a thermopower along the temperature gradient with a magnitude depending on the transverse spin current boundary conditions. This SMT can be distinguished from the conventional Seebeck effect when modulating the transverse spin current boundary conditions by the spin transfer torque (STT) at the ferromagnetic insulator/normal metal (FMI/N) interface \cite{Ralph:spintorque:JMMM:2008, Non-collinear-spintronics:Brataas:2006}.
The STT depends on the orientation of the magnetization $\mathbf{M}$ in the magnetic insulator. When $\mathbf s$ and $\mathbf{M}$ are collinear (either parallel or antiparallel), the STT vanishes. This situation corresponds to open transverse spin transport boundary conditions [cf. Fig.\,\ref {fig2}(e)].
In contrast, when $\mathbf s$ and $\mathbf{M}$ enclose a finite angle, the STT is finite, becoming maximal for $\mathbf s$ orthogonal to $\mathbf{M}$ [short-circuit boundary conditions, Fig.\,\ref {fig2}(f)].
We control the transverse spin current boundary conditions by systematically changing the orientation of the magnetization in the FMI layer, and record the ensuing spin Nernst driven changes in the thermopower, i.e., the SMT.
The phenomenology of the SMT is similar to the recently established spin Hall magnetoresistance (SMR) \cite{Nakayama:SMR:PRL:2013, SMR:torque:Vlietstra:APL2013}.\\
We model the SMT, by the spin diffusion equation in the metal with quantum mechanical boundary conditions at the ferromagnet, as detailed in the SI.
The (longitudinal) thermopower $V_{\mathrm{thermal}}$ can be expressed in terms of an effective Seebeck coefficient as:
\begin{equation}
\frac{V_{\mathrm{thermal}}}{l} = -[S +\Delta S_0 +\Delta S_1 (1-m_{\mathrm{y}}^2)]\partial_x T
\label{E1}
\end{equation}
with
\begin{equation}
\frac{\Delta S_1}{S} = -\theta_{\mathrm{SN}}\theta_{\mathrm{SH}}\frac{\lambda}{t_{\mathrm{N}}}\mathrm{Re}\frac{2\lambda G \tanh^2(\frac{t_{\mathrm{N}}}{2\lambda})}{\sigma+2\lambda G \coth (\frac{t_{\mathrm{N}}}{\lambda})}.
\label{E2}
\end{equation}
Here, $t_\mathrm{N}$, $\sigma$ and $\lambda$ are the thickness, electrical conductivity, and spin diffusion length of the N film, respectively, $G$ the spin mixing conductance of the FMI/N interface, $l$ the sample length and $m_{\mathrm y}=\mathbf{M}\cdot\mathbf{y} /|\mathbf{M}|$. The $t_{\mathrm{N}}$-dependence of $\Delta S_1/S$ in Eq.\,(\ref{E2}) is identical to that of the SMR \cite{Chen:SMR:theory:PRB:2013}.

\begin{figure}[hp]
  \centering
  \includegraphics[width=\textwidth]{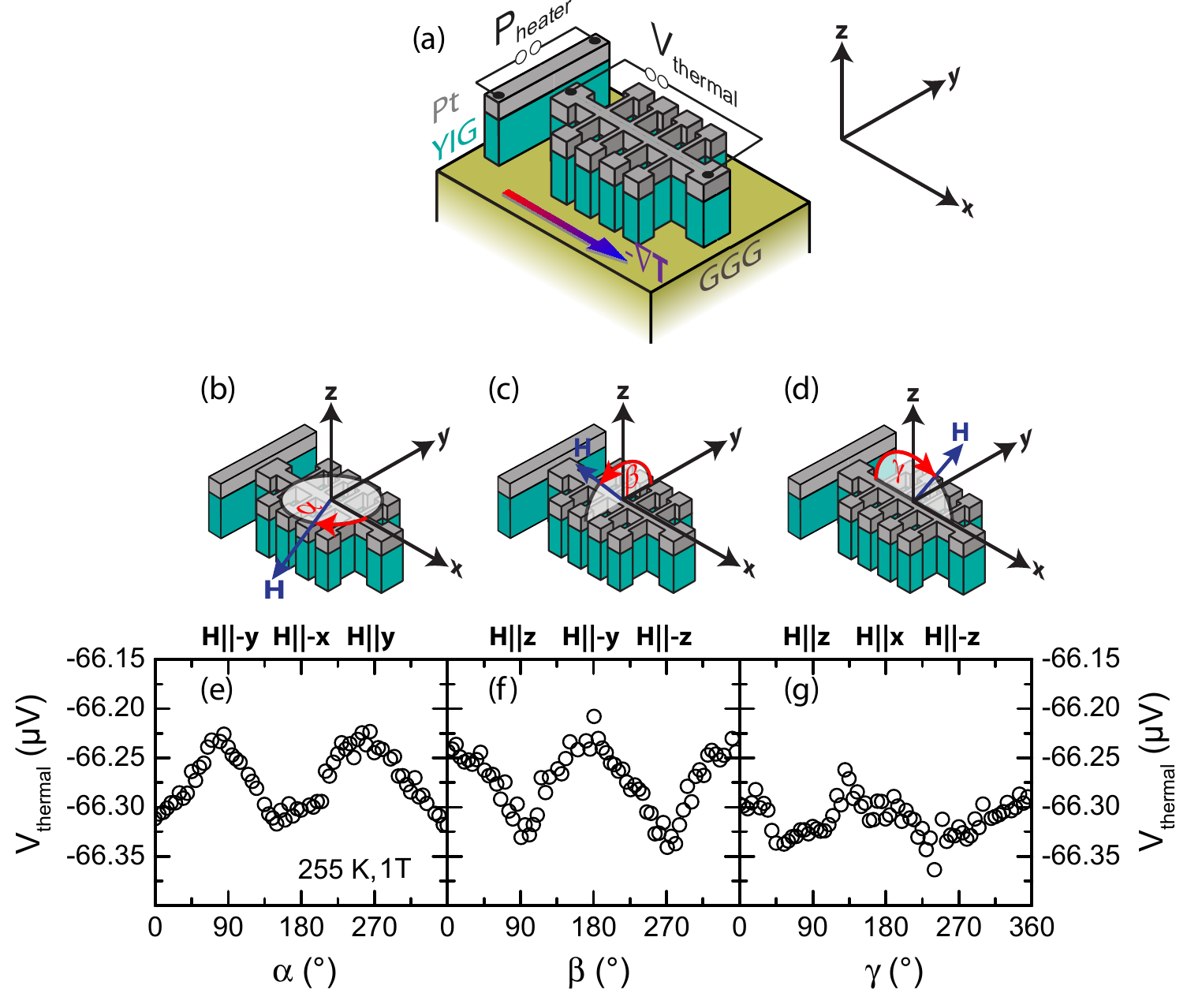}
  \caption{\textbf{(a)} Setup of the SMT experiments. A YIG$|$Pt ($t_\mathrm{F}=40\,\mathrm{nm}/t_\mathrm{N}=4.1\,\mathrm{nm}$) thin film is patterned into a Hall bar (width $w=250\,\mu\mathrm m$, length $l=3150\,\mu\mathrm m$). An additional heater strip  is defined along $\mathbf y$, $d=250\,\mu\mathrm m$ beyond the top of  the Hall bar.
  By applying an electric current with power $P_\mathrm{heater}$ to the heater strip, one end of the Hall bar is hotter than the other end that is connected to a heat sink provided by the sample holder (see supplementary information for details), leading to a temperature gradient $-\nabla T$ along $\mathbf x$.
  \textbf{(b)-(d)} The magnetization vector $\mathbf M$ of the YIG layer is rotated by an external magnetic field $\mu_0 H = 1\,\mathrm T$ in three different rotation planes spanned by ($\mathbf x$, $\mathbf y$) (panel \textbf{(b)}), ($\mathbf y$, $\mathbf z$)(panel \textbf{(c)}) and ($\mathbf x$, $\mathbf z$)(panel \textbf{(d)}). The measured thermal voltage $V_\mathrm{thermal}$ for all three geometries and $P_\mathrm{heater} = 286\,\mathrm{mW}$ (or $\Delta T = 18.0\,\mathrm K$ along the Hall bar, corresponding to $T_\mathrm{sample}\approx 255\,\mathrm{K}$) is depicted in panels \textbf{(e)} for the ($\mathbf x$, $\mathbf y$)-plane, \textbf{(f)} for the ($\mathbf y$, $\mathbf z$)-plane and \textbf{(g)} for the ($\mathbf x$, $\mathbf z$)-plane.}
	\label{fig3}
\end{figure}

The sample is a yttrium iron garnet ($\mathrm{Y}_3\mathrm{Fe}_5\mathrm{O}_{12}$, YIG)$|$Pt bilayer \cite{Althammer:SMR:experiment:PRB:2013}
patterned into a Hall bar as shown in Fig. \ref{fig3}(a). An additional YIG$|$Pt strip extending in $\mathbf{y}$ direction serves as an on-chip heater.
We heat one side of the sample by applying a constant electric power of $286\,\mathrm{mW}$
to the on-chip heater and connect the other end of the sample to a heat sink. This generates a temperature difference $ \Delta T = T_{\mathrm {hot}} - T_{\mathrm {cold}} = 18.0\,\mathrm{K}$ between the two ends of the Hall bar as measured by on-chip resistive thermometry (see SI), while the dip stick is kept at $T_\mathrm{base}=220\,\mathrm{K}$; the average sample temperature for these heater settings amounts to $T_\mathrm{sample}\approx 255\,\mathrm{K}$(see SI).\\
The external magnetic field of $\mu_0 H = 1\,\mathrm{T}$ is much larger than the demagnetization and anisotropy fields of YIG, such that $\mathbf{M} \parallel \mathbf{H}$. Then, $\mathbf{H} || \mathbf{y}$ corresponds to $\mathbf{M} || \mathbf{s}$ and thus open boundary conditions (no spin current flow across the interface), while for $\mathbf{H} || \mathbf{x}$ and $\mathbf{H} || \mathbf{z}$ the ferrimagnet represents an efficient spin current sink resulting in maximum spin current flow across the interface.
The thermopower $V_{\mathrm{thermal}} = (\tilde{S}+\Delta S_1) \Delta T$ measured along $\mathbf x$ [cf. Fig. \ref{fig3}(b)] contains the conventional Seebeck effect of the Pt Hall bar with the Seebeck coefficient $\tilde{S}$ (for details see SI). For Pt, $\tilde{S}<0$, such that the corresponding thermopower is negative.
By rotating $\mathbf H$ and therefore the magnetization $\mathbf M$ of the YIG within the film plane from $\alpha = 0^\circ$ ($\mathbf H \parallel \mathbf x$) to $\alpha = 90^\circ (\mathbf H \parallel \mathbf y)$, the spin current boundary conditions are switched from short-circuit (finite transverse spin current) to open circuit (vanishing transverse spin current) conditions. The thermopower therefore shows a characteristic modulation as expressed by Eq.\,(\ref{E1}).\\
Our measurements confirm this expectation: For open boundary conditions, $V_{\mathrm{thermal}} = -66.225\,\mu\mathrm{V}$ is about $\Delta V_{\mathrm{thermal}} = 100\,\mathrm{nV}$ larger than for short-circuit conditions, with a relative signal amplitude of $\left| \Delta V_{\mathrm{thermal}}/V_{\mathrm{thermal}} \right|= (1.5\pm 0.3)\times 10^{-3}$, see Fig. \ref{fig3}(e).
We reproduced this behavior for full $360^\circ$ rotations of the applied field in the sample plane spanned by $\mathbf x$ and $\mathbf y$, leading to a $\sin^2 \alpha$ behavior of $V_{\mathrm{thermal}}$ with minima for short-circuit boundary conditions ($\alpha = 0^\circ, 180^\circ$), and maxima for open boundary conditions ($\alpha = 90^\circ, 270^\circ$).
We can also switch the boundary conditions by rotating the magnetic field in the (normal) plane spanned by $\mathbf y$ and $\mathbf z$, see Fig.\,\ref{fig3}(f).
Starting at $\beta = 0^\circ$ from $\mathbf H \parallel\mathbf y$ (open boundary conditions), the thermal voltage decreases while rotating $\mathbf H$ towards $\mathbf z$ ($\beta = 90^\circ$, short-circuit boundary conditions) and the minimal and maximal levels of $V_{\mathrm{thermal}}$ coincide with the ones obtained in the first geometry.
Rotating $\mathbf H$ out-of-plane perpendicular to $\mathbf y$ [Fig. \ref{fig3}(g)], the signal is almost constant and coincides with the lower signal levels observed for the other rotation planes. This is exactly the behavior expected from Eq.\,(\ref{E1}), since $\mathbf H \perp \mathbf s$ is fulfilled for every magnetization orientation in this rotation geometry, causing a maximum spin Nernst spin current flow.
Also the observed transverse thermopower agrees very well with theory (see SI, Fig.\,6). Spurious effects can be ruled out by their symmetries. For example, a spin Seebeck voltage arising from $\nabla T$ along $\mathbf{z}$ would result in a $\sin(\alpha)$ [$\cos(\beta)$] dependence of $V_{\mathrm{thermal}}$ in the ($\mathbf x$,$\mathbf y$) [($\mathbf y$,$\mathbf z$)] rotation plane, which is not observed (see also SI). Control experiments conducted on a GGG/Pt sample exhibit no SMT signature (see SI).
Using $\lambda = 1.5\,\mathrm{nm}$, $\theta_\mathrm{SH}=0.11$ and $\mathrm{Re}(G) = 4\times 10^{14}\,\Omega^{-1}\mathrm{m}^{-2}$ \cite{Meyer:SMR:T-dep:APL:2014} in Eq.\,(\ref{E2}), the observed $\Delta V_{\mathrm{thermal}}/V_{\mathrm{thermal}} = -1.5\times 10^{-3}$ corresponds to a spin Nernst angle of $\theta_{\mathrm{SN}} = -0.20$ for Pt. Our first-principles calculations for the spin transport in bulk Pt confirm the relative sign and suggest $\theta_{\mathrm{SH}}/\theta_{\mathrm{SN}} = -0.6$ at $T_\mathrm{sample}$(see SI). This is in fair agreement with $\theta_{\mathrm{SH}}/\theta_{\mathrm{SN}} (\mathrm{exp}.) = -0.5$.
For different heating powers between $100\,\mathrm{mW}$ and $286\,\mathrm{mW}$ as well as for two different magnetic field values $\mu_0H=0.5\,\mathrm{T}$ and $1\,\mathrm{T}$, we obtain identical SNE signatures.
The relative amplitude of the modulation of the thermal voltage does not depend on both heating power and external magnetic field, as expected (see SI).
Note that the observed field dependence excludes interference by the spin Seebeck effect \cite{Caloritronix:SpinSeebeck:Uchida:Saitoh:Nature:2008, Caloritronix:SpinSeebeck:GaMnAs:Jaworski:Myers:NatMat:2010, Longitudinal-spin-Seebeck:Uchida:APL:2010:172505}.\\
In summary, we report an SMT signal in Pt$|$YIG hybrids proportional to an in-plane temperature gradient that reveals the spin Nernst effect in Pt, thereby opening a new strategy for the thermal generation of spin currents.
The spin Nernst and spin Hall angles are of equal magnitude in Pt but of opposite sign, as corroborated by first principle calculations.
With the observation of the spin Nernst effect, the ``zoo" of magneto-thermo-galvanic effects is complete.
We anticipate that this spin Nernst magneto-thermopower can help in the optimization of spintronic devices harvesting ubiquitous temperature gradients e.g. from Joule heating hot spots. \textit{Note added}: While writing this manuscript, we became aware of an additional experiment on the spin Nernst effect in metallic multilayers.\,\cite{Sheng2016}
\newpage
\section{Methods summary}
In our experiments, an yttrium iron garnet ($\mathrm{Y}_3\mathrm{Fe}_5\mathrm{O}_{12}$, YIG) thin film was epitaxially grown on a (111)-oriented gadolinium gallium garnet ($\mathrm{Gd}_3\mathrm{Ga}_5\mathrm{O}_{12}$, GGG) substrate by pulsed laser deposition, covered in-situ with Pt by electron beam evaporation \cite{Althammer:SMR:experiment:PRB:2013}.
The YIG film is a an insulating ferrimagnet with a saturation magnetization of $120\,\mathrm{kA/m}$.
The Pt layer is polycrystalline with a resistivity of $430\,\mathrm{n}\Omega\mathrm{m}$ at room temperature.
The thicknesses of the YIG and Pt layers were determined by x-ray reflectometry as $t_{\mathrm{F}}= (40\pm 2)\,\mathrm{nm}$ and  $t_{\mathrm{N}}= (4.1\pm 0.2)\,\mathrm{nm}$, respectively.
The $5\times5\,\mathrm{mm}^2$ sample is patterned into a Hall bar with an additional heating strip as shown in Fig.\,\ref{fig3}(a).
For temperature differences $\Delta T \leq 18.0\,\mathrm{K}$ between both ends of the Hall bar, the voltage signal $V_{\mathrm{thermal}}$ is measured while rotating a magnetic field of constant magnitude $\mu_0 H = 1\mathrm{T}$ in different planes.
$\mu_0 H$ is much larger than the saturation field of YIG to ensure the alignment of the magnetization $\mathbf M$ of the FMI parallel to the external field even in the presence of magnetic and shape anisotropies. More details on the experimental methods are given in the SI.

\section{Acknowledgments}
S.M., M.A., S.G., H.H., R.G. and S.T.B.G. thank Andreas Erb for the preparation of the stoichiometric YIG target, Thomas Brenninger for technical support, and Nynke Vlietstra for the non-local sample preparation.
Y.-T.C. and G.E.W.B. acknowledge funding by the FOM (Stichting voor Fundamenteel Onderzoek der Materie), EU- ICT-7 ``INSPIN", and Grant-in-Aid for Scientific Research (Grant Nos. 25247056, 25220910, 26103006). S.W. thanks Sebastian Tölle and Ulrich Eckern for helpful discussions.
All authors acknowledge funding via the DFG Priority program 1538 ``Spin-Caloric Transport" (Projects GO 944/4, BA 2954/2 and EB 154/25).
\section{Author Contributions}
S.M., R.S., T.W. designed the sample layout and carried out the experiments. S.M., Y.-T.C., G.E.W.B., R.G. and S.T.B.G. developed the explanation of the SMT effect. S.G. supervised the sample growth. Y.-T.C. and G.E.W.B. developed the theoretical framework and S.W., D.K. and H.E. performed first-principles calculations of the relevant spin-caloric transport coefficients.
S.T.B.G. supervised the experiments. The manuscript was written by S.M., M.A. and S.T.B.G. All authors discussed and participated in writing the manuscript under the guidance of S.M. and G.E.W.B.

\section{Affiliations}
\textbf{Walther-Mei{\ss}ner-Institut, Bayerische Akademie der Wissenschaften, Walther-Mei{\ss}ner-Stra{\ss}e 8, 85748 Garching, Germany}\\
Sibylle Meyer, Matthias Althammer, Stephan Gepr\"{a}gs, Hans Huebl, Rudolf Gross and Sebastian T. B. Goennenwein\\
\textbf{Physik-Department, Technische Universit\"at M\"unchen, 85748 Garching, Germany}\\
Sibylle Meyer, Hans Huebl, Rudolf Gross and Sebastian T.B. Goennenwein
\textbf{Kavli Institute of NanoScience, Delft University of Technology, Lorentzweg 1, 2628 CJ Delft, The Netherlands}\\
Yan-Ting Chen and Gerrit E. W. Bauer\\
\textbf{RIKEN Center for Emergent Matter Science (CEMS), 2-1 Hirosawa, Wako, Saitama 351-0198, Japan}\\
Yan-Ting Chen\\
\textbf{Department Chemie, Physikalische Chemie, Universit\"at M\"unchen, Butenandtstra{\ss}e 5-13, 81377 M\"unchen, Germany}\\
Sebastian Wimmer, Diemo K\"odderitzsch and Hubert Ebert\\
\textbf{Nanosystems Initiative Munich (NIM), Schellingstra{\ss}e 4, 80799 M\"unchen, Germany}\\
Hans Huebl, Rudolf Gross and Sebastian T. B. Goennenwein\\
\textbf{Institute for Materials Research, Tohoku University, Sendai, Miyagi 980-8577, Japan}\\
Gerrit E. W. Bauer\\
\textbf{WPI Advanced Institute for Materials Research, Tohoku University, Sendai 980-8577, Japan}\\
Gerrit E. W. Bauer\\

\section{Competing Interests Statement}
The authors declare that they have no competing financial interests. Supplementary information accompanies
this paper online.
\section{Corresponding author}
Correspondence and requests for materials should be addressed to S.T.B.G.

\newpage

\end{document}